\documentclass[manuscript]{acmart}

\AtBeginDocument{%
  }

\copyrightyear{2026}
\acmYear{2026}
\setcopyright{rightsretained}
\acmConference[LAK '26]{Learning Analytics and Knowledge Conference 2026}{April 27-May 1, 2026}{Bergen, Norway}
\acmBooktitle{Learning Analytics and Knowledge Conference 2026 (LAK '26), April 27-May 1, 2026, Bergen, Norway}\acmDOI{10.XXXX}

\acmISBN{XXX-X-XXXX-XXXX-X/2026/04}

\begin{document}

\title{Productive Discussion Moves in Groups Addressing Controversial Issues}


\author{Kyuwon Kim}
\affiliation{%
  \institution{Ewha Womans University}
  \city{Seoul}
  \country{Republic of Korea}}
\email{kyuwonkim95@ewha.ac.kr}

\author{Jeanhee Lee}
\affiliation{%
  \institution{Ewha Womans University}
  \city{Seoul}
  \country{Republic of Korea}}
\email{jinnylee26@ewha.ac.kr}

\author{Sung-Eun Kim}
\affiliation{%
  \institution{Ewha Womans University}
  \city{Seoul}
  \country{Republic of Korea}}
\email{sungeunkim@ewhain.net}

\author{Hyo-Jeong So}
\affiliation{%
  \institution{Ewha Womans University}
  \city{Seoul}
  \country{Republic of Korea}}
\email{hyojeongso@ewha.ac.kr}
\renewcommand{\shortauthors}{Kim et al.}

\begin{abstract}
 Engaging learners in dialogue around controversial issues is essential for examining diverse values and perspectives in pluralistic societies. While prior research has identified productive discussion moves mainly in STEM-oriented contexts, less is known about what constitutes productive discussion in ethical and value-laden discussions. This study investigates productive discussion in AI ethics dilemmas using a dialogue-centric learning analytics approach. We analyze small-group discussions among undergraduate students through a hybrid method that integrates expert-informed coding with data-driven topic modeling. This process identifies 14 discussion moves across five categories, including Elaborating Ideas, Position Taking, Reasoning \& Justifications, Emotional Expression, and Discussion Management. We then examine how these moves relate to discussion quality and analyze sequential interaction patterns using Ordered Network Analysis. Results indicate that emotive and experiential arguments and explicit acknowledgment of ambiguity are strong positive predictors of discussion quality, whereas building on ideas is negatively associated. Ordered Network Analysis further reveals that productive discussions are characterized by interactional patterns that connect emotional expressions to evidence-based reasoning. These findings suggest that productive ethical discussion is grounded not only in reasoning and justification but also in the constructive integration of emotional expression.
\end{abstract}


\begin{center}
\large
\textcolor{red}{Accepted as a Full Research Paper at the}\\
\textcolor{red}{16th International Conference on Learning Analytics \& Knowledge (LAK 2026)}

\end{center}

\maketitle

\section{Introduction}
 As society grows increasingly diverse due to rapid technological advances, there is an urgent need to address controversial issues that remain underexplored. Controversial issues, spanning from socio-scientific topics to political and historical debates, are commonly defined as those involving conflicting explanations held by diverse stakeholder groups \cite{santos2025critical}. These issues often draw upon ethical principles and underlying premises that vary across individuals, thereby provoking fundamental disagreements \cite{levinson2006towards, great1998education, oulton2004reconceptualizing}. The complexity of such controversies is further intensified by technological advances, which introduce a broader range of stakeholders \cite{purcell2025estimating, rudschies2020value}. As stakeholders approach these issues with differing beliefs, understandings, and moral values \cite{clarke2005learning, oulton2004reconceptualizing}, learners need to be equipped with the capacity to critically examine and engage with diverse worldviews \cite{geldof2018superdiversity}.

Engaging in dialogue is a promising way for learners to encounter diverse perspectives \cite{kim2025your}. Despite growing interest in dialogue as a pedagogical approach, the definition of dialogue remains contested within the field of dialogic education \cite{calcagni2018three, howe2017commentary, alexander2019whose}. To provide conceptual clarity, this study adopts Alexander's conceptualization, which distinguishes\textit{ dialogue }from more general exchanges based on its purposeful structure and educational intent \cite{kim2019dialogic}. While \textit{discussion} typically refers to a relatively open exchange of ideas, \textit{dialogue} is defined here as a deliberately facilitated form of interaction aimed at shared understanding \cite{alexander2001conversation, alexander2019whose}. This study focuses on \textit{discussion moves}, defined as individual utterances that serve distinct interactional functions. By adopting discussion moves as the primary unit of analysis, the study examines how these exchanges contribute to \textit{productive discussion}, conceptualized as dialogue in which multiple perspectives are actively engaged with and integrated through interaction.

Beyond serving as a mere exchange of ideas, dialogue has strong cognitive potential to support learners in reflecting on their own positions while engaging with alternative viewpoints on controversial issues \cite{kim2025your, wang2025flip}. Fostering such dialogue is therefore essential for promoting deeper engagement and productive discussion \cite{alexander2001conversation}. In practice, however, classroom discussions often fall short of this ideal. Although productive dialogue depends on exposure to disagreement and the integration of diverse perspectives \cite{hennessy2023analysis, kinchin2016charting}, group dynamics frequently undermine these processes, leading learners to remain silent, converge prematurely on superficial consensus, or defer to dominant voices \cite{chinn2000structure, gillies2019promoting, smith1981can}. These challenges highlight the need for scaffolding and guidance to support dialogic engagement.

Providing such support is difficult in classroom settings where teachers must manage multiple groups simultaneously \cite{michaels2012talk}. Consequently, prior research has explored technology-based supports, including collaborative conversational agents \cite{de2025investigating, nguyen2023role}, and instructional dashboards \cite{gutierrez2023analytical, tegos2019towards}. Learning analytics plays a central role in enabling these approaches by identifying interaction patterns and supporting automated interventions in learners’ dialogic processes \cite{de2024learning, banihashem2022systematic}. For such interventions to be effective, it is essential to establish a clear understanding of what constitutes a productive conversation.

Prior studies have sought to identify productive discussion moves in classroom settings \cite{alexander2001conversation, alexander2017towards}. Most of the work, however, has focused on scientific understanding and reasoning \cite{santos2025critical, asterhan2013epistemic}, resulting in a lack of frameworks that adequately capture the nature of ethical discussions \cite{nucci2016recovering, brandel2024dialogue}. To fully realize the potential of dialogue around ethical and controversial issues, it is therefore necessary to examine what counts as productive discussion in such contexts. In this study, we focus on ethical dilemmas related to AI, which expose tensions between human values and technological possibilities \cite{li2023norms, solovyeva2018going}. Placing learners in dilemmatic situations can evoke both intra- and interpersonal conflict, which plays a central role in shaping discussion dynamics \cite{kim2025dilemmas}. Building on this potential, the current study addresses the following research questions:
\begin{itemize}
    \item RQ1. What types of discussion moves emerge when learners engage with controversial issues?
    \item RQ2. Which discussion moves are associated with higher- quality discussions?
    \item RQ3. How do the sequential patterns of discussion differ between high- and low- quality discussion groups?
\end{itemize}

\section{Related Works}

\subsection{The Role of Dialogue in Addressing Controversial Issues}

One defining feature of contemporary society is the coexistence of diverse values and perspectives \cite{tessler2024ai}. Pluralistic societies create conditions in which differing interpretations and claims can coexist regarding the same issue. Consequently, topics that require social consensus inevitably take on the character of controversial issues, where opposing positions collide \cite{levinson2006towards}. Controversial issues are approached by a wide range of individuals and groups who hold conflicting premises, beliefs, and understandings. As such, each position may be logically reasoned but cannot be resolved solely through scientific evidence \cite{great1998education, oulton2004reconceptualizing, wellington1985including}. Because controversial issues are shaped by stakeholders who approach them from diverse beliefs, understandings, and moral values, education must foster learners’ capacity to explore and critically reflect on multiple worldviews \cite{clarke2005learning}.

In this context, dialogic education has gained increasing attention as a pedagogical approach that exposes learners to diverse perspectives and provides a means to address polarization \cite{erstad2024education, kolikant2025constructive}. Dialogue, characterized as purposeful and considerate discussion on a shared topic, positively influences traditional aspects of learning performance, including subject knowledge, comprehension, and reasoning skills \cite{resnick2015introduction}. Prior research has further demonstrated that learners who engage in dialogue with peers holding different viewpoints on a task tend to achieve higher performance outcomes \cite{doise1979individual, mugny1978socio, muhonen2018quality}.

More recently, dialogue has been recognized as a tool for mediating polarization in society  \cite{santos2025critical}. Its strong cognitive potential is particularly crucial in the context of sensitive and controversial issues, encouraging learners to reflect on their own positions while exploring alternative viewpoints on controversial issues \cite{wang2025flip}. When diverse perspectives collide, such conflicts create opportunities for learners to critically examine competing viewpoints, reconsider underlying assumptions, and refine their judgments on complex issues \cite{haidt2001emotional}. Therefore, it is crucial to provide learners with an open and inclusive dialogic space that supports meaningful engagement across different perspectives \cite{alexander2019whose}.

\subsection{Discussion Moves as Indicators of Productive Discussion}

As the importance of dialogue has become widely recognized, a body of research has sought to determine when discussion can be considered productive. Two well-known approaches are  \textit{Accountable Talk} \cite{michaels2008deliberative, resnick2018accountable}, which emphasize mutual accountability toward dialogue among learners \cite{alexander2010speaking}, and \textit{Exploratory Talk} \cite{mercer2008value, mercer2002exploratory}, which highlights critical and constructive engagement \cite{mercer2002exploratory}. Building on these dialogic frameworks, prior work has identified empirical indicators of productive discussion and specific moves associated with discussion quality. A recent meta-analysis reported 28 forms of discussion moves shown to positively affect interaction quality, domain-specific learning, and problem-solving performance \cite{hu2024systematic, michaels2015conceptualizing}. Although scholars vary in how they conceptualize productive peer interaction \cite{howe2017commentary, kim2019dialogic}, a common understanding is that specific discussion moves foster dialogic productivity.

Because most prior studies have focused on reasoning and argumentation in scientific contexts, a significant gap remains in frameworks that adequately capture the nature of ethical and value-laden discussions \cite{nucci2016recovering, brandel2024dialogue, asterhan2013epistemic}. In response to this limitation, recent efforts have begun to explore productive discussion in ethical domains. For example, Brandel et al. \cite{brandel2024dialogue} introduced the \textit{Dialogue on Ethics and Ethics of Dialogue} framework, which offers valuable insights into ethical reasoning during classroom discussions. Yet, because this framework operates at a holistic level rather than at the granularity of individual utterances, its applicability for identifying specific discussion moves remains limited. Similarly, Felton et al. \cite{felton2022capturing} proposed analytic schemes for deliberative argument related to reasoning and persuasion. However, this work primarily emphasizes argumentative and persuasive processes, leaving aspects of value-laden dialogue less specified. Building on these studies, our research seeks to extend dialogue-centric learning analytics to capture the distinctive features of controversial discussion and to identify productive discussion moves groups addressing AI ethics dilemmas.

\subsection{Hybrid Approach for Dialogue-Centric Learning Analytics}

To analyze conversational data, it is essential to code the transcription with discussion moves to uncover meaningful patterns. Traditionally, most studies have employed a top-down approach, relying on expert-developed speech taxonomies as a critical step \cite{traum200020}. While top-down approaches are effective for evaluating the effectiveness of a designed system, they have several limitations. First, they heavily depend on the individuals who design and apply speech taxonomies, making it difficult for different annotators and even machines to reliably assign the same labels to similar utterances \cite{traum200020}. Second, expert-generated taxonomies often fail to capture insights that emerge outside predefined categories, as they are typically developed without a systematic and exhaustive analysis of the available data \cite{andernach1997finding}.

To address these limitations, initial attempts were made to automatically discover speech acts through bottom-up approaches. For example, Rus et al. \cite{rus2012automated} proposed a data-driven method that applied K-Means clustering based on leading tokens of utterances. Similarly, Ezen-Can et al. \cite{ezen2013unsupervised} introduced an unsupervised approach to classify discussion patterns without relying on predefined taxonomies. These works represent important initial steps toward hybrid approaches.  The key challenge, as highlighted in prior research, lies in finding a balance between extrinsic forces (expert knowledge) and intrinsic forces (data-driven discovery). Such a balance could lead to a more robust taxonomy of discussion moves, which is simultaneously informed by expert perspectives and grounded in empirical data \cite{rus2012automated, xu2020applying}. This vision resonates with the ultimate goal of Dialogue-Centric Learning Analytics (DCLA): to contribute to theory building and to “close the loop” in learning analytics research \cite{knight2015discourse}.

In this work, we propose language models as a promising solution for the hybrid approach. Recent developments in language modeling allow artificial intelligence systems to capture the semantic meaning of sentences by processing sequential data \cite{li2020sentence}. With this potential, many researchers have leveraged sentence embeddings to reduce the labor-intensive process of manual coding and to complement expert-defined taxonomies in dialogue analysis \cite{lin2023enhancing, lin2023robust, duran2023sentence, yin2025scaling}. These advances suggest that language models can play a central role in realizing hybrid approaches that combine expert knowledge with data-driven discovery.

\section{Methods}
\subsection{Participants and Contexts}
We recruited 51 undergraduate students from diverse majors to participate in group discussions on controversial issues. Participants were randomly assigned to groups of three. After completing a pre-survey on demographic information, each group engaged in a face-to-face discussion session focused on AI ethics dilemmas, particularly scenarios involving human-like household robots. To highlight tensions between human values and technological possibilities, the researchers designed five dilemmatic scenarios, summarized in Table \ref{tab:scenario}.

\begin{table}[htbp]
\centering
\caption{Short Descriptions of Dilemmatic Scenarios}
\resizebox{\textwidth}{!}{%
\begin{tabular}{l}
\toprule
  \textbf{Descriptions of Dilemmatic Scenarios} \\
\midrule
 1. A household android requires 24-hour private data collection for a system upgrade. Should the user allow or restrict it?\\
 2. After an emotional upgrade, a user’s mother became attached to a household android. Should further emotional upgrades be allowed?\\
 3. Parents increasingly rely on household androids to interact with children. Should the country impose age restrictions?\\
 4. Decisions without clear explanations raises calls for XAI despite performance and security concerns. Should XAI be mandated?\\
 5. Household androids are essential but environmentally costly. Should their upgrade and use be restricted?\\
\bottomrule
\end{tabular}
}

\label{tab:scenario}
\end{table}

Before the experiment began, participants received instructions about the experiment procedure. Each session lasted approximately 50-60 minutes, during which groups discussed the dilemmas sequentially and were instructed to reach a joint consensus at the end. The whole process was recorded and later transcribed for analysis. Each participant received 50,000 KRW (approximately USD 36) as compensation. We collected a total of 85 audio recordings (5 sessions × 17 groups), each lasting 10-20 minutes. Two sessions affected by technical errors or premature interruptions were excluded, resulting in 83 sessions included in the final analysis.

\subsection{Data Analysis}
Our study aimed to explore a hybrid analytic approach that integrates expert knowledge with data-driven discovery to examine discussion moves associated with productive discussion about controversial issues. The study proceeded in two phases. First, to identify specific discussion moves that emerge in ethical discussions, we developed and implemented a hybrid analytic approach.\footnote{For the code used in this study, see: https://github.com/kyuw0nkim/LAK26-methodology} In line with prior work on hybrid analysis \cite{xu2020applying}, we began with a deductive phase based on pre-defined schemes and progressed toward data-driven discovery of interactional structure. For the top-down coding, we applied the Scheme for Educational Dialogue Analysis (SEDA), originally developed by Hennessy et al. \cite{hennessy2016developing}. SEDA is one of the most widely used coding schemes for systematically analyzing productive discussion across diverse educational contexts \cite{tao2023coding}. The scheme was selected for this study because it is designed to capture diverse dialogue types (e.g., \textit{Invite elaboration or reasoning}, \textit{Make reasoning explicit}, \textit{Build on ideas}, \textit{Express or invite ideas}, \textit{Positioning and coordination}, \textit{Reflect on the dialogue or activity}, \textit{Connect}, and \textit{Guide direction of dialogue or activity}), rather than focusing solely on a limited set of predefined productive moves. Following the eight major SEDA clusters, a total of 2,199 utterances were independently coded by two annotators. To establish reliability, the annotators independently coded 20\% of the data, achieving a Cohen’s Kappa of .8, which indicates substantial agreement. Any discrepancies that emerged during the annotating process were resolved through discussion among the researchers.

Next, we employed BERTopic to identify latent topics corresponding to discussion moves through a clustering approach. Clustering is an unsupervised method that groups data points based on similarity. BERTopic leverages clustering techniques to generate latent topic representations \cite{grootendorst2022bertopic}. Although it is typically used for document-level topic discovery, recent studies have demonstrated its potential for classifying dialogue acts using BERT embeddings \cite{zhang2024predicting}. We used the Korean Sentence-BERT model \cite{gan2021kosroberta} to embed each utterance. Because the semantic similarity of dialogue topics makes it difficult for the model to differentiate utterances in the embedding space, we fine-tuned the model on 20\% of our manually labeled data. This fine-tuning guided the model to represent different discussion moves more distinctly. The fine-tuned Sentence-BERT was used to generate contextual embeddings for all utterances, which were subsequently clustered using HDBSCAN. To control granularity, we set the minimum topic size to 15 and the topic merging threshold to 0.5.

After clustering, each discovered topic was reviewed by researchers for post-hoc interpretation. To assess whether clusters aligned with SEDA, we applied the LIFT score of expert-assigned labels. The LIFT score quantifies how much more likely a label is to occur within a cluster than in the dataset overall, defined as $LIFT(label, clsuter) = P(label |cluster)/P(label)$.
We adopted the LIFT score instead of simple proportions because the high frequency of certain dominant labels could otherwise skew interpretation. In this study, we set a LIFT threshold of 2.0, meaning that a label was at least twice as likely to appear within a cluster compared to its overall frequency, and treated such labels as strongly characterizing the cluster. Based on this criterion, we considered clusters containing a single label exceeding the LIFT threshold were interpreted in conjunction with the original coding scheme. Two types of discussion moves not captured by the taxonomy also emerged: (1) clusters with no labels exceeding 2.0, and (2) clusters with multiple labels exceeding 2.0, indicating that the original scheme could not adequately distinguish between them.  We considered these patterns as novel discussion moves identified in our study.

With the discussion moves identified in RQ1, we then conducted dialogue analysis to address RQ2. Specifically, we employed a linear mixed-effects model to examine the relationship between discussion moves and discussion quality. This analysis accounts for the nested structure of the data in which multiple conversation sessions (up to five) were derived from the same group. Discussion quality was operationalized using Integrative Complexity (IC) scores, proposed by Suedfeld et al. \cite{suedfeld199227}. Reflecting the extent to which groups considered and integrated diverse perspectives \cite{baker1990coding, jennstaal2019deliberation}, IC provides a useful measure for assessing the quality of group discussions on complex and value-laden topics \cite{park2018effects, brodbeck2021group}. For reliability, three researchers independently evaluated 24\% of the sessions (20 sessions) referring to the IC coding scheme \cite{baker1990coding} and resolved disagreements through discussion. The initial inter-rater agreement was \textit{r}\textsubscript{wg} = .73, ICC (1) = .47, and ICC (2) = .72. To further enhance reliability, the researchers jointly assessed an additional 20\% of the remaining data. The final agreement reached \textit{r}\textsubscript{wg} = .94, ICC (1) = .60, and ICC (2) = .75.

Lastly, we performed Ordered Network Analysis (ONA) to examine dialogic patterns during ethical discussion (\textit{RQ3}). ONA is a data mining technique widely used to investigate transition patterns among student actions and collaborative processes \cite{tan2022ordered}. In this study, we compared groups with high and low discussion quality to gain insights into productive discussions on controversial issues. Specifically, groups in the top quartile (upper 25\% based on IC scores) were classified as high-quality discussion groups, while those in the bottom quartile (lower 25\%) were classified as low-quality discussion groups.

\section{Findings}
\subsection{Types of Discussion Moves in Controversial Issues}

\begin{figure}
    \centering
    \includegraphics[width=0.5\linewidth]{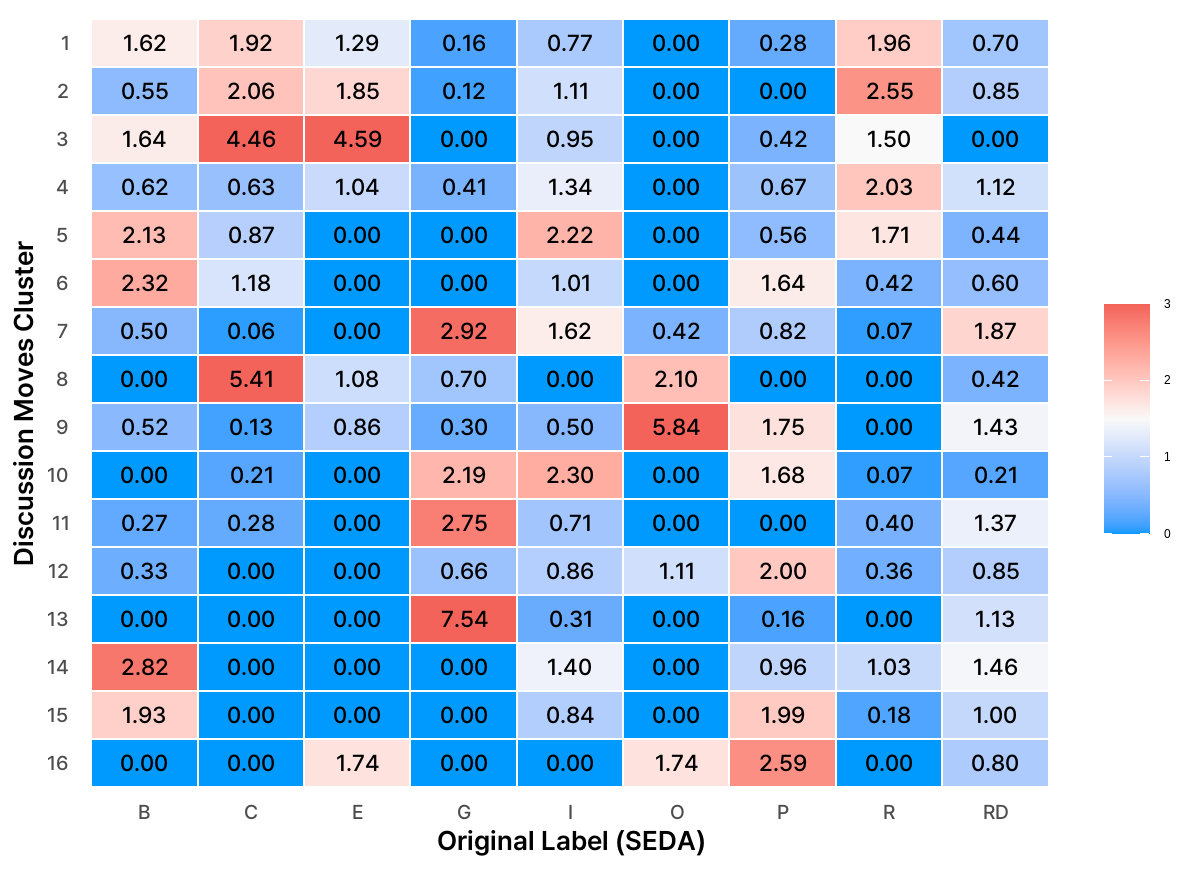}
    \caption{Alignment between discovered discussion move clusters and SEDA labels}
    \label{fig:heatmap}
\end{figure}

We performed a bottom-up analysis to identify the distinct types of discussion moves that emerged during learners' discussions on AI ethics dilemmas (RQ1). Using a clustering approach with BERTopic, we initially obtained 16 clusters. To evaluate whether these clusters aligned with existing dialogue taxonomies, we applied the LIFT score of expert-assigned labels. The LIFT score indicates how strongly a label characterizes a cluster. We visualized the alignment between clusters and labels through a heatmap (see Fig ~\ref{fig:heatmap}). Clusters with no dominant label or with multiple overlapping labels were considered novel discussion moves discovered in our work.

Among the 16 initial clusters, seven represented new discussion types that were not adequately captured by existing SEDA labels. After merging conceptually similar clusters, we identified a final set of \textit{14 distinct discussion moves}, which were further organized into five broader categories: (1) Elaborating Ideas, (2) Position Taking, (3) Reasoning \& Justifications, (4) Emotional Expression, and (5) Discussion Management. This data-driven process, supported by LIFT-based validation and post-hoc human review, enabled us to develop a comprehensive coding scheme that captures the nuances of ethical conversation. The complete scheme, including descriptions and examples for each discussion move, is presented in Table ~\ref{tab:talkmoves}.

The first category, \textit{Elaborating Ideas}, encompasses moves that deepen and expand the content of the discussion. This includes \textit{Perspective Taking}, where participants explicitly state their stance or role; \textit{Building on Ideas}, where they summarize or extend a peer’s contribution. Those moves represent a finer-grained differentiation of the \textit{Building on Ideas} construct in the SEDA scheme, reflecting prior work that highlights examining issues from multiple perspectives as a key feature of socio-scientific reasoning \cite{sadler2007students}. \textit{Connecting with External Contexts} also emerged as an independent move, consistent with findings that real-world connections are used to clarify complex problems and elicit cognitive dissonance in dialogue \cite{wang2025flip}.

The second category, \textit{Position Taking}, relates to how participants articulate and navigate their stances within the group. This category features moves such as \textit{Inviting Perspectives}, where a learner actively solicits other stances; \textit{Position Declaration}, which marks a clear and final statement of one’s decision \cite{felton2022capturing}; and \textit{Certainty Expression}, where participants convey the level of confidence or uncertainty they feel about their position. This reflects the uncertain feature of the learner dealing with an ethical dilemma \cite{ames2016moral}.

The third and most central category of argumentation is \textit{Reasoning \& Justifications}. These moves constitute the logical core of discussions on controversial issues, in which learners articulate claims through explicit reasoning and maintain a skeptical stance toward others  \cite{driver2000establishing, sadler2007students}. Within this category, we identified \textit{Logical Justification}, which involves using causal or conditional reasoning to support a claim; \textit{Evidence-based Argument}, where participants draw on factual knowledge, data, or policies as evidence; and \textit{Critical Questioning}, a move used to challenge or probe another participant’s reasoning.

In addition to cognitive and logical moves, we found that participants frequently used moves related to \textit{Emotional Expression}. This category highlights the value-laden nature of ethical dilemmas \cite{sadler2004morality}. It includes \textit{ Emotive/Experiential Argument}, where personal feelings or lived experiences are used to support a stance, and \textit{Acknowledging Ambiguity}, where participants openly express their internal conflict or the difficulty of the dilemma. In particular, this category represents a discussion move that has newly emerged in the context of ethical discussions, underscoring the distinctive characteristics of moral and value-based reasoning.

Finally, the \textit{Discussion Management} category includes meta-level discussion moves that structure and guide the conversation itself. These moves are in line with social metacognitive activities used during the group discussion \cite{molenaar2014metacognitive}. These are functional moves, such as \textit{Procedure Management}, which involve suggestions on how to proceed with the discussion; \textit{Facilitating Agreement}, used to explicitly check for group consensus; and \textit{Conversation Maintenance}, which includes minimal responses like "Right" or "uh-huh" that serve to maintain conversational flow and signal active listening. These identified discussion moves serve as the foundational variables for our subsequent analyses, enabling us to investigate their relationship with the overall quality of group discussion (RQ2) and uncover specific discussion patterns (RQ3).

\begin{table}[ht]
\centering
\caption{Coding scheme of discussion moves in ethical discussions. Novel discussion moves identified through our analysis are marked with (*)}

\begin{tabular}{p{1.5cm}p{2cm}p{5cm}p{5cm}}\toprule

\textbf{Categories}& \textbf{Labels}& \textbf{Descriptions}&\textbf{Examples} \\\midrule

Elaborating Ideas& Perspective Taking& Elaborating on one's own role, stance, or perspective to make an idea more detailed and explicit.   &"I'm speaking from the perspective of an environmental group representative right now." \\

& Building on Ideas& Summarizing, rephrasing, or extending another participant’s idea to deepen the discussion.   &"So, you're saying a robot needs to act like a robot." \\

& Connecting with External Contexts *& Connecting the current issue to external cases (e.g., daily life, history, technology, policy) to explain or justify one’s stance.   &"Think about when laptops were first developed; their energy efficiency was terrible at the start... So maybe this robot is just in that initial phase." \\\midrule
 Position Taking& Inviting Perspectives& Asking about another participant’s stance or the reasons behind it.   
&"How about you? Do you agree with the statement?" \\
 & Position Declaration *& Clearly stating one’s final decision or position at the end of a discussion.   
&"I absolutely think restrictions are needed." \\
 & Certainty Expression *& Expressing certainty or uncertainty regarding one’s stance or choice.   &"I agreed, but I'm still not sure about it..." \\\midrule
 Reasoning \& Justifications& Logical Justification  & Justifying a stance using logical structures such as causality or conditional reasoning.   
&"One of the basic consumer rights is the right to be informed. So this is a right. That's why I think it's necessary." \\

& Evidence-based Argument *& Using factual knowledge, policies, or technical data as evidence to support one’s argument.   
&"Recently, there's been a growing number of policies to introduce explainable AI... and that's why I think it should be mandatory." \\

& Critical Questioning *& Asking probing questions aimed at challenging or rebutting another participant’s reasoning.   &"So, as a corporation, you won't make disclosure mandatory?" \\\midrule

Emotional Expression& Emotive/ Experiential Argument *& Supporting one’s stance with personal feelings, emotions, or lived experiences.   
&"I'm an only child... and honestly, as someone who always wanted siblings, I don't think this robot is such a bad idea." \\

& Acknowledging Ambiguity *& Expressing internal conflict, difficulty, or confusion when facing a complex dilemma.   &"This is tough." / "This is a hard one." \\\midrule

Discussion Management& Procedure Management& Suggesting ways to manage or structure the discussion process.&"Should we discuss this again?" \\
 & Facilitating Agreement& Collecting group opinions explicitly through polls, votes, or (dis)agreement checks.&"Let's collect our opinions." \\

& Conversation Maintenance *& Providing minimal utterances (e.g., “yes”, “uh-huh”) to maintain conversational flow.&"Right." / "Exactly." \\ \bottomrule

\end{tabular}
\label{tab:talkmoves}
\end{table}

\subsection{Impact of Discussion Moves on Discussion Quality}

We performed a statistical analysis using a linear mixed-effect model to examine which discussion moves significantly predicted the quality of group discussion, as measured by Integrative Complexity. The frequencies of 14 discussion moves (see Table ~\ref{tab:talkmoves})  served as independent variables. The full results of the analysis are presented in Table \ref{tab:sorted_estimates}.

The results revealed that four specific discussion moves were statistically significant predictors of discussion quality. In particular, moves within the \textit{Emotional Expression} category showed the strongest positive associations. \textit{Emotive/Experiental Argument} emerged as the most significant positive predictor (B = 0.081, \textit{p} = .005), suggesting that grounding arguments in personal feelings and lived experiences meaningfully enhanced the quality of group discussions. In addition, \textit{Acknowledging Ambiguity} was identified as a significant predictor (B = 0.436, \textit{p} = .040), indicating that groups that explicitly recognized the inherent complexity and uncertainty of ethical dilemmas tended to engage in more sophisticated and nuanced discussions. In contrast, there was one discussion move that demonstrated a significant negative relationship with discussion quality. Within the \textit{Elaborating Ideas} category, \textit{Building on Ideas} was negatively associated with discussion quality  (B = -0.283, \textit{p} = .040). From the \textit{Reasoning \& Justifications} category, \textit{Evidence-based Argument} emerged as a marginally positive predictor of discussion quality (B = 0.171, \textit{p} = .058), indicating a trend whereby the use of factual knowledge, policy references, or technical data may contribute to higher-quality ethical discussions. Other discussion moves, including those related to logical justification, perspective-taking, and discussion management, did not exhibit statistically significant associations with discussion quality in the model.

\begin{table}[h]
    \centering
        \caption{Results of the linear mixed-effects model predicting discussion quality from discussion moves. ( · \textit{ p} < .1, *\textit{ p} < .05, **\textit{ p} < .001)}
    \resizebox{\textwidth}{!}{%
    \begin{tabular}{llrrrrl}
        \hline
         && \textbf{Estimate (B)}& \textbf{Std. Error}& \textbf{\textit{t} value}& \textbf{\textit{p} value}& \\ 
        \hline
         Elaborating Ideas&Connecting with External Contexts& 0.164& 0.111& 1.478& .144&  
\\ 
         &Perspective Taking& 0.064& 0.142& 0.448& .656&  
\\ 
         &Building on Ideas& -0.283& 0.136& -2.090& .040& *
\\ 
        \hline
         Position Taking&Position Declaration& 0.156& 0.192& 0.813& .419&  
\\ 
         &Inviting Perspectives& -0.076& 0.072& -1.038& .303&  
\\ 
         &Certainty Expression& 0.124& 0.178& 0.694& .490&  
\\ 
        \hline
         Reasoning \& Justifications&Evidence-based Argument& 0.171& 0.089& 1.932& .058& ·\\
         &Logical Justification& 0.012& 0.111& 0.106& .916&  
\\
         &Critical Questioning& -0.012& 0.164& -0.071& .944&  
\\ 
        \hline
         Emotional Expression&Emotive/Experiential Argument& 0.081& 0.028& 2.941& .005& **
\\ 
         &Acknowledging Ambiguity& 0.436& 0.197& 2.088 & .040& * \\ 
        \hline
         Discussion Management&Procedure Management& 0.034& 0.055& 0.606& .547&  
\\
         &Facilitating Agreement& -0.138& 0.125& -1.101& .275&  
\\ 
         &Conversation Maintenance& -0.038& 0.044& -0.875& .385&  
\\ 
        \hline
         &(Intercept)& 2.342& 0.247& 9.502& 1.9e-10& 
\\ 
        \hline
 \textbf{Random Effects \& SD}& Group (Intercept)& 0.28
& & & &\\
 & Residual& 0.93
& & & &\\
    \hline
    \end{tabular}
    }
    \label{tab:sorted_estimates}
\end{table}

\subsection{Ordered Network of Discussion Moves in Controversial Contexts}

\begin{figure}
    \centering
    \includegraphics[width=0.85\linewidth]{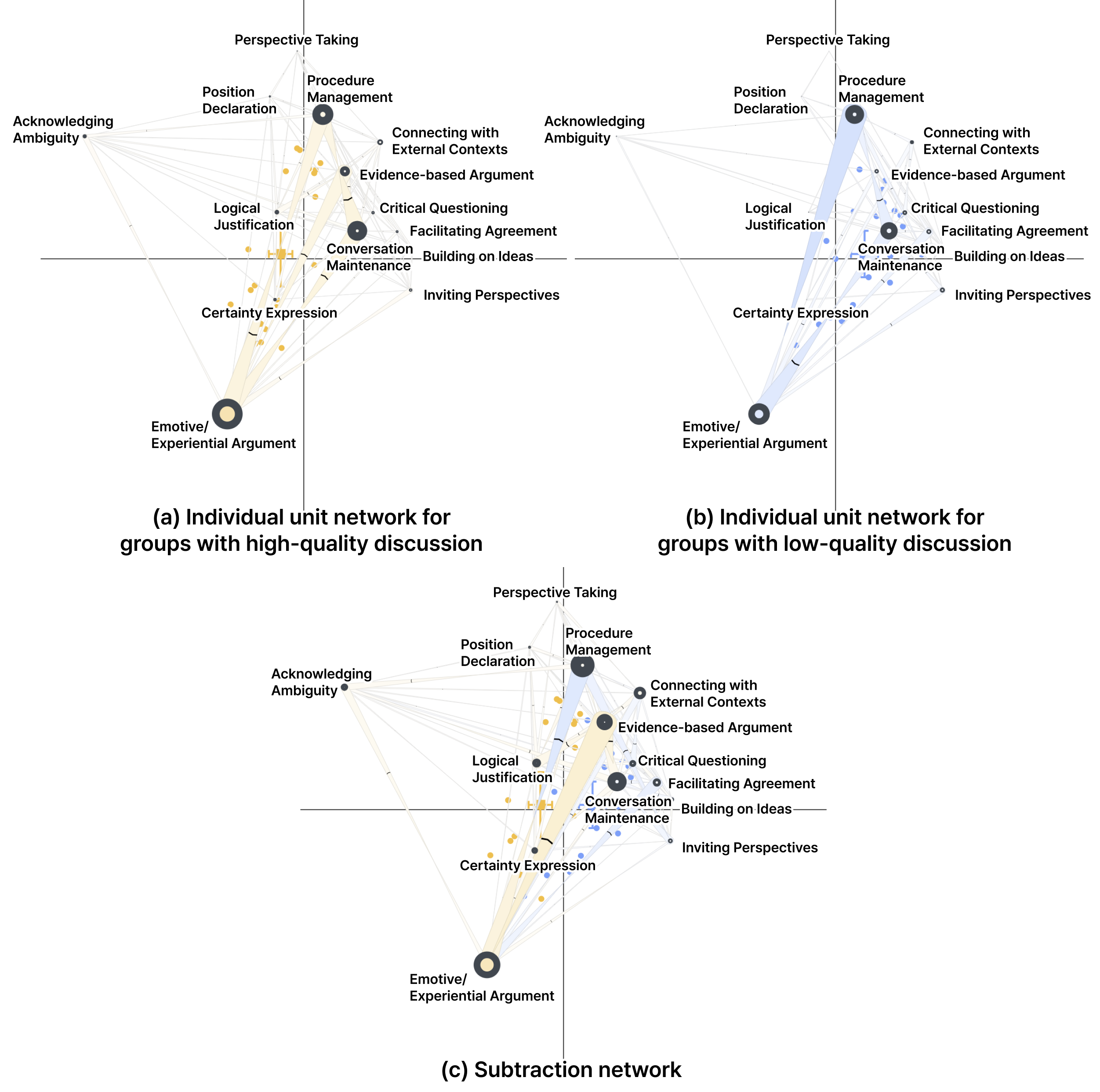}
\caption{Ordered network of discussion moves in different discussion quality}
    \label{fig:ona}
\end{figure}

We conducted ONA to identify and compare the structured network and sequential patterns of discussion moves in groups with high and low discussion quality. Discussion-level specific networks were constructed by identifying groups in the first (Q1) and fourth (Q4) quartiles based on the IC scores of the discussion data. High-quality discussion groups had a mean score of 4.61, whereas low-quality groups had a mean score of 1.70.  Fig \ref{fig:ona}a, Fig \ref{fig:ona}b visualizes the individual unit network for interaction patterns in discussion quality. Fig \ref{fig:ona}c presents the subtraction network between high- and low-quality discussion groups, representing the differences in their interaction patterns. Differences are indicated using distinct colors, with interaction patterns that are more strongly exhibited in each group represented using the corresponding color.

As shown in Fig \ref{fig:ona}a, the most prominent node in the high-quality network is \textit{Emotive/Experiential Argument}, followed by discussion managing behaviors such as \textit{Procedure Management} and \textit{Conversational Maintenance}. Focusing on the direction of the arrows, \textit{Procedure Management} was followed by \textit{Emotive/Experiential Argument}, leading to \textit{Conversational Maintenance}. The largest inner circle in \textit{Emotive/Experiential Argument} indicates that learners in high-quality discussion groups repeatedly interact with others through emotional and experiential arguments rather than logical arguments.

Within low-quality discussion groups (see Fig \ref{fig:ona}b), the most prominent nodes were \textit{Emotive/Experiential Argument}, \textit{Procedure Management}, and \textit{Conversational Maintenance}, like those observed in the high-quality groups. However, the \textit{Emotive/Experiential} Argument was not strongly pronounced. Particularly, the directionality of edges differed from that of the high-quality groups, with prominent arrows running from \textit{Emotive/Experiential Argument} to \textit{Procedure Management }and from \textit{Conversational Maintenance} to \textit{Emotive/Experiential Argument}. In the low-quality discussion group, \textit{Emotive/Experiential Argument} also exhibited the most frequent self-transitions among all discussion moves.

Fig \ref{fig:ona}c visualizes the differences between the two groups by highlighting the transition patterns that characterize each discussion type. Beyond the previously noted differences in arrow directionality, the high-quality discussion group exhibits a strong transition from \textit{Emotive/Experiential Argument} to \textit{Evidence-based Argument}, represented by a prominent edge in the network.

\section{Discussion}

This study aimed to unpack the nature of productive discussion in the context of ethical deliberation around controversial issues. By employing a hybrid, data-driven analytic approach, we identified key discussion moves (RQ1), examined their relationship with discussion quality (RQ2), and analyzed the sequential patterns that differentiate high- and low-quality discussions (RQ3). Together, these findings advance our understanding of productive discussion in value-laden contexts and inform implications for practice and future research.

The hybrid approach for developing coding schemes (RQ1) yielded a nuanced framework of 14 discussion moves, categorized into five broad themes. Through this process, we contribute a hybrid analytic approach for identifying discussion moves in dialogue, enabling the systematic integration of theory-driven and data-driven perspectives. While the resulting framework included expected categories such as \textit{Reasoning \& Justifications}, the hybrid process also revealed interactional dimensions that are often underrepresented in pre-defined schemes. Particularly, \textit{Emotional Expression}, including \textit{Emotive/Experiential Argument} and \textit{Acknowledging Ambiguity}, emerged as a salient theme through the bottom-up analysis. This finding highlights the limitation of discussion analysis frameworks often developed in STEM contexts, which tend to prioritize cognitive and logical discussion moves \cite{alexander2017towards}. Discussions around controversial issues, however, inherently involve ethical deliberation. As Johnson \cite{johnson2015morality} notes, such a dialogic process is simultaneously emotional, rational, and imaginative. Our results empirically demonstrate that in ethical discussions addressing controversial issues, emotional and affective expressions are not incidental but are integral components of the discussion. This aligns with prior works that emphasize the role of emotions, values, and personal experiences in navigating controversial issues \cite{nucci2016recovering}.

The importance of affective moves in discussion about controversial issues was reinforced by our linear mixed-effect model analysis (RQ2). The finding that mentioning emotional and personal experience was a powerful positive predictor of high-quality discussion challenges the conventional view that emotional appeals are less valuable than logical reasoning \cite{foster2013teaching}. In the context of ethical dilemmas with no clear right-or-wrong solution, personal experiences and feelings serve as a valid form of evidence, helping participants to empathize with different perspectives and explore the human impact of their decisions. Similarly, the positive impact of acknowledging ambiguity suggests that high-quality ethical discussion is not about finding a single correct answer, but about grappling with complexity. This pushes prior work by Gaydos et al. \cite{gaydos2025game}, which highlighted the emergence of moral ambiguity as students explore ethical issues, by empirically demonstrating that such ambiguity can be productive in compelling deeper discussion. Groups that acknowledged and openly admitted the ambiguity of the problem engaged in more sophisticated and integrative discussions. Evidence-based argument marginally predicted a higher quality of discussion. This confirms that grounding claims in factual data remains a cornerstone of productive discussion, even in value-laden topics.

The most counterintuitive finding, however, was the negative association between \textit{Building on Ideas} and the discussion quality. One possible explanation is that while this move appears collaborative on the surface, it may have often manifested as simple rephrasing or superficial agreement (e.g., “So you’re saying... yes, I agree”). Prior research has shown that uncritical and affirmative repetition or extension of others’ contributions is less productive than genuinely building on each other’s ideas, as it can lead to superficial consensus rather than substantive engagement \cite{knight2015role, mercer1996quality}. Its negative correlation in our study may indicate that participants engaged in building on ideas through affirmative loops to avoid deeper conflict or critical engagement, thereby settling for surface-level consensus rather than pushing the dialogue forward \cite{chinn2000structure}. This finding indicates that the discussion moves aimed at building on ideas appear to hinge on subtle distinctions between productive elaboration and unproductive affirmation. Future research should investigate what linguistic or contextual features allow language models to reliably differentiate these cases of dialogic productivity.

ONA approach (RQ3) provided a narrative for how dialogic moves combine to create productive or unproductive discussions. The key difference between high- and low-quality discussion groups was not merely the frequency of certain moves, but the interactive patterns they formed. In high-quality discussions, emotive or experiential arguments were not treated as endpoints. Instead, they were repeatedly revisited and elaborated upon, often through self-transitions, and subsequently connected to evidence-based arguments. This pattern suggests that personal or emotional expressions served as resources for advancing ethical reasoning. Importantly, these transitions were embedded within sequences that sustained the flow of discussion, as procedural management and conversational maintenance moves supported continued exploration and extension of ideas. In contrast, low-quality discussions exhibited a different interactional configuration. Although emotive or experiential arguments were also present, they were more likely to transition into procedural management or to recur following conversational maintenance moves. In these cases, emotional or experiential expressions tended to function as requests for structure or as repeated affective responses, rather than as contributions that advanced the line of reasoning. Transitions from emotive or experiential arguments to evidence-based arguments were comparatively weak, and discussions often stalled without progressing toward more analytically grounded exchanges.

These findings resonate with prior qualitative research on dialogue around controversial issues. Firer et al. \cite{firer2021quality} demonstrated that differences between groups that regulated emotions and those that did not were primarily reflected in cognitively and argumentatively oriented moves, such as \textit{Reasoning Clarification} and \textit{Building on Ideas}. Our findings extend this work by revealing similar patterns through an analytic, network-based approach, showing how such differences emerge from patterns of discussion moves. While emotions can foster engagement and involvement in dialogue, they need to be interactively controlled and nuanced to support productive deliberation \cite{englund2016moral}. Building on our results, we suggest that one form of such control may lie in facilitating transitions that connect emotive or experiential expressions to subsequent reasoning processes. Taken together, our findings indicate that discussion quality is shaped by whether emotive and experiential contributions are interactionally positioned to support the development of more reasoned and integrative arguments. Emotional expression alone does not distinguish high- from low-quality discussions. Rather, it is the sequential organization through which such expressions are elaborated and connected to evidence-based reasoning that differentiates productive discussion from interactional stagnation.

\subsection{Implications for Practice}

This study provides implications for how productive discussions may be supported in practice. Rather than focusing solely on the presence of specific dialogue moves, our results highlight the importance of attending to how contributions are positioned and taken up in relation to one another over time. Just as successful problem-solving groups actively engaged with and built upon one another’s contributions, thereby establishing a joint problem-solving space \cite{barron2003smart}, our findings suggest that the uptake and temporal interconnectedness of contributions are also critical for productive ethical discussions. The effectiveness of dialogue moves, therefore, depended not on their occurrence alone, but on whether and how they were connected to subsequent responses and reasoning. From this perspective, a key role of learning analytics and AI-supported technologies lies in helping learners and teachers notice or support these interactional relationships as discussions unfold. This could take the form of dialogue dashboards that reflect temporal patterns of interaction (e.g., \cite{wang2025enhancing}), or AI-based approaches that leverage cognition modeling to capture evolving contextual information within dialogue (e.g., \cite{samadi2024ai, kim2025your}). By making visible how contributions are taken up, elaborated, or left unaddressed over time in discussions, learning analytics can support more informed facilitation and reflection without prescribing specific forms of participation.

\section{Conclusion}
In an increasingly diverse and technologically complex society, learners need to engage in dialogue about controversial issues that involve conflicting values and perspectives. This study contributes to that need by identifying discussion moves that arise in ethical discussions, examining how they relate to group discussion quality, and analyzing the sequential patterns that distinguish high- from low-quality dialogue. Our findings show that productive discussions are not only grounded in reasoning and justification but also enriched by emotional expression, lived experiences, and openness to ambiguity. By highlighting these characteristics, findings reported in this study can be used to design scaffolds and analytics that can foster reflective and productive discussions in pluralistic classrooms.

Further research is needed to explore the potential of hybrid analytic approaches and to examine productivity in ethical discussions. Some limitations of this study should be noted. First, since the participants in this study were Korean undergraduate students, the findings may not be generalizable to learner profiles with different cultural and language contexts. Because ethical judgment is closely intertwined with cultural nuances \cite{zheng2014influence}, future research is needed to validate the proposed analytic approach with broader and more heterogeneous samples. Second, this study relied on manual annotation to provide theoretically grounded supervision for a transformer-based model, as no open-sourced datasets labeled according to the SEDA framework were available. Consequently, the researchers directly coded the dialogue data to establish analytic ground truth, which limits the scalability of the approach. Future research could explore the development of shared datasets or pre-trained models to reduce manual effort while preserving analytic fidelity. Lastly, to derive meaningful clusters of dialogue, parameters such as the minimum topic size and the topic merging threshold were manually determined by the researchers. Although this tuning was necessary to ensure interpretable results, it introduces a degree of subjectivity. Future work should therefore explore more systematic or data-driven approaches to parameter selection to enhance robustness and reproducibility. 

\begin{acks}
This work was supported by the National Research Foundation of Korea (NRF) grant funded by the Korea government (MSIT) (RS-2024-00350045).
\end{acks}

\bibliographystyle{ACM-Reference-Format}
\bibliography{LAK_references}

\end{document}